\documentclass[aps,prd,superscriptaddress,showpacs,preprint,amsmath,amssymb]{revtex4}
\usepackage{graphicx, bm}
\usepackage[usenames]{color}

\begin{document}

\draft
\title{Studies on the anomalous magnetic and electric dipole moments of the tau-neutrino in $pp$ collisions at the LHC}

\author{ A. Guti\'errez-Rodr\'{\i}guez\footnote{alexgu@fisica.uaz.edu.mx}}
\affiliation{\small Facultad de F\'{\i}sica, Universidad Aut\'onoma de Zacatecas\\
         Apartado Postal C-580, 98060 Zacatecas, M\'exico.\\}

\author{M. K\"{o}ksal\footnote{mkoksal@cumhuriyet.edu.tr}}
\affiliation{\small Deparment of Optical Engineering, Cumhuriyet University, 58140, Sivas, Turkey.\\}

\author{A. A. Billur\footnote{abillur@cumhuriyet.edu.tr}}
\affiliation{\small Deparment of Physics, Cumhuriyet University, 58140, Sivas, Turkey.\\}

\author{ M. A. Hern\'andez-Ru\'{\i}z\footnote{mahernan@uaz.edu.mx}}
\affiliation{\small Unidad Acad\'emica de Ciencias Qu\'{\i}micas, Universidad Aut\'onoma de Zacatecas\\
         Apartado Postal C-585, 98060 Zacatecas, M\'exico.\\}

\date{\today}

\begin{abstract}

In this paper the production cross section $pp\rightarrow (\gamma, Z) \to \nu_\tau \bar \nu_\tau \gamma+X$ in $pp$ collisions at
$\sqrt{s}=8, 13, 14, 33\hspace{0.8mm}TeV$ is presented. Furthermore, we estimate bounds at the $95\%\hspace{1mm}C. L.$ on the dipole
moments of the tau-neutrino using integrated luminosity of ${\cal L}=20, 50, 100, 200, 500, 1000, 3000 \hspace{0.8mm}fb^{-1}$ collected
with the ATLAS detector at the LHC and we consider systematic uncertainties of $\delta_{sys}=0, 5, 10\hspace{1mm}\%$. It is shown that
the process under consideration is a good prospect for probing the dipole moments of the tau-neutrino at the LHC.

\end{abstract}


\vspace{5mm}

\maketitle

\section{Introduction}

Due to remarkable advances in science and technology of accelerators \cite{ATLAS5,CMS5}, many experiments on elementary
particle physics are being carried out to test the predictions of the Standard Model (SM) \cite{Glashow,Weinberg,Salam}.
Many of these experiments are being performed at the present time in the Large Hadron Collider (LHC) at the TeV scale.

The physics program of the ATLAS collaboration at the LHC \cite{Aad} contemplates the study of the hadroproduction of
$Z$ bosons associated with one or two photons. To carry out your study the ATLAS collaboration use ${\cal L}=20.3\hspace{0.8mm}fb^{-1}$
of $pp$ collisions collected with the detector operating at a center-of-mass energy of $\sqrt{s}=8\hspace{0.8mm}TeV$. For their analysis
they use the decays $Z/\gamma^* \rightarrow l^+ l^-$ with $l =(e^- \hspace{0.8mm}\mbox{or} \hspace{0.8mm}\mu)$ and $Z \to \nu \bar \nu$.
The production channels studies are $pp\rightarrow l^+l^-\gamma + X$ and $pp\rightarrow l^+l^-\gamma \gamma + X$. Another important
channels are $pp\rightarrow \nu \bar \nu \gamma + X$ and $pp\rightarrow \nu \bar \nu \gamma \gamma + X$. In all the production channels,
the study is made with no restriction on the recoil system $X$ (inclusive events) and by requiring that the system $X$ have
no central jet (exclusive events).

A fundamental challenge of the particle physics community is to determine the Majorana or Dirac nature of the neutrino.
For respond to this challenge, experimentalist are exploring different reactions where the Majorana nature may manifest
\cite{Zralek}. About this topic, the study of neutrino magnetic moments is, in principle, a way to distinguish between
Dirac and Majorana neutrinos since the Majorana neutrinos can only have flavor changing, transition magnetic moments
while the Dirac neutrinos can only have flavor conserving one.

Another interesting topic contemplated by the ATLAS collaboration, is the study of the dipole
moments of the neutrino through the processes $pp\rightarrow (\gamma, Z) \to \nu \bar \nu \gamma + X$ and
$pp\rightarrow (\gamma, Z)\to \nu \bar \nu \gamma \gamma + X$, which have not been explored before in hadron colliders.
In this regard, one of the most active fields in high-energy physics is the theoretical and experimental investigation
of neutrino properties as well as of their interactions. The study of the physics of the neutrino is a powerful
tool and contribute us precious information on the physics of the SM and provides a window on the new physics beyond the SM.

In Refs. \cite{Sahin,Sahin1} the electromagnetic properties of the neutrinos were examined at the LHC via the processes
$pp \rightarrow p \gamma^{*} p\rightarrow p \nu \bar{\nu} q X$ and $pp \rightarrow p\gamma^{*} \gamma^{*} p\rightarrow p \nu \bar{\nu} p$,
respectively.

In the SM the neutrinos are massless particles of which theirs electromagnetic property
are poorly known experimentally. In addition, the observation of neutrino oscillation shows the necessity of neutrino masses, which implies
that the SM to be modified such that non-trivial electromagnetic structure of neutrino should be reconsidered \cite{Giunti}. In the minimal
extension of the SM to incorporate the neutrino mass the anomalous magnetic moment (MM) of the neutrino is known to be developed in one loop calculation,
$\mu_\nu=\frac{3eG_F m_{\nu_i}}{(8\sqrt{2}\pi^2)}\simeq 3.1\times 10^{-19}(\frac{m_{\nu_i}}{1 \hspace{1mm} eV})\mu_B$, where $\mu_B=\frac{e}{2m_e}$
is the Bohr magneton \cite{Fujikawa,Shrock}, and the non-zero mass of the neutrino is essential to get a non-vanishing magnetic moment. Furthermore,
the SM predict CP violation, which is necessary for the existence of the electric dipole moments (EDM) of a variety physical systems. The EDM provides
a direct experimental probe of CP violation \cite{Christenson,Abe,Aaij}, a feature of the SM and beyond SM physics. The signs of new physics can be
analyzed by investigating the electromagnetic dipole moments of the tau neutrino, such as its MM and EDM.

Limits on the dipole moments of the tau-neutrino have been reported by different experiments at Borexino \cite{Borexino}, E872 (DONUT) \cite{DONUT},
CERN-WA-066 \cite{A.M.Cooper}, and at LEP \cite{L3}, these are summarized in Table I. Another limits
on the MM and EDM of the tau-neutrino in different context are reported in the literature \cite{Gutierrez12, Gutierrez11,Gutierrez10,Gutierrez9,Gutierrez8,Data2016,Gutierrez7,Sahin,Sahin1,Gutierrez6,Aydin,Gutierrez5,Gutierrez4,Gutierrez3,Keiichi,Aytekin,Gutierrez2,
Gutierrez1,DELPHI,Escribano,Gould,Grotch}.

\begin{table}[!ht]
\caption{Summary of limits on the dipole moments of the tau-neutrino.}
\begin{center}
\begin{tabular}{|c| c| c| c|}
\hline
{\bf Collaboration}  &    {\bf Experimental limit}                                   & {\bf C. L.}  &  {\bf Reference}\\
\hline
\hline
Borexino       &   $\mu_{\nu_\tau} < 1.9\times 10^{-10}\mu_B$ & $90 \%$  & \cite{Borexino} \\
\hline
E872 (DONUT)   &   $\mu_{\nu_\tau} < 3.9\times 10^{-7}\mu_B$  & $90 \%$  & \cite{DONUT} \\
\hline
CERN-WA-066    &   $\mu_{\nu_\tau} < 5.4\times 10^{-7}\mu_B$   & $90 \%$ & \cite{A.M.Cooper} \\
\hline
L3             &   $\mu_{\nu_\tau} < 3.3\times 10^{-6}\mu_B$   & $90 \%$  & \cite{L3} \\
\hline
\hline
{\bf Model}                       &     {\bf Theoretical limit}                             &    {\bf C. L.}  &  {\bf Reference}\\
\hline
\hline
Vector-like multiplets     &     $d_{\nu_\tau} < O( 10^{-18}-10^{-20} \hspace{0.8mm}e cm)$  & $95 \%$  & \cite{Tarek} \\
\hline
Model independent           &     $d_{\nu_\tau} < O(2 \times 10^{-17} \hspace{0.8mm}e cm)$   & $95 \%$  & \cite{Keiichi} \\
\hline
Effective Lagrange approach &     $d_{\nu_\tau} < 5.2\times 10^{-17}\hspace{0.8mm}e cm$      & $95 \%$  & \cite{Escribano} \\
\hline\hline
\end{tabular}
\end{center}
\end{table}

Motivated for the physical program of the ATLAS collaboration with regard to the study on the dipole moments of the neutrino, in this paper we
explore the possibility of probing the dipole moments of the tau-neutrino through the process $pp\rightarrow (\gamma, Z) \to \nu_\tau \bar \nu_\tau \gamma + X$ for $\sqrt{s}=8, 13, 14, 33\hspace{0.8mm}TeV$. We use integrated luminosity of ${\cal L}=20, 50, 100, 200, 500, 1000, 3000\hspace{0.8mm}fb^{-1}$ collected with the ATLAS detector at the LHC and we consider systematic uncertainties of $\delta_{sys}=0, 5, 10\hspace{1mm}\%$. All our study was carried out with a $95\%$ confidence level (C. L.). It is shown that the process under consideration is a good prospect for probing the dipole moments of the tau-neutrino at the LHC.

This paper is structured as follows. In Section II, we study the total cross section and the dipole moments of the tau-neutrino through the
channel $pp\rightarrow (\gamma, Z) \to \nu_\tau \bar \nu_\tau \gamma + X$. Section III provides the conclusions.

\section{Production process $\nu_\tau\bar\nu_\tau\gamma$ in $pp$ collisions}

The electromagnetic properties of any fermion appear in quantum field theory, through  its interaction with the photon. In the present paper,
we study the electromagnetic properties of a neutral fermion, such as the neutrino. For this purpose, we following a focusing as the performed
in our previous works \cite{Gutierrez12,Gutierrez11,Gutierrez10,Gutierrez9,Gutierrez8,Gutierrez7,Gutierrez6,Gutierrez5,Gutierrez4,Gutierrez3,Gutierrez2,Gutierrez1}. Therefore, the most general expression for the vertex of interaction $\nu_\tau\bar\nu_\tau\gamma$ is given by

\begin{equation}
\Gamma^{\alpha}=eF_{1}(q^{2})\gamma^{\alpha}+\frac{ie}{2m_{\nu_\tau}}F_{2}(q^{2})\sigma^{\alpha
\mu}q_{\mu}+eF_3(q^2)\gamma_5\sigma^{\alpha\mu}q_\mu +eF_4(q^2)\gamma_5(\gamma^\mu q^2-q\llap{/}q^\mu),
\end{equation}

\noindent where $e$ is the charge of the electron, $m_{\nu_\tau}$ is the mass of the tau-neutrino, $q^\mu$
is the photon momentum, and $F_{1, 2, 3, 4}(q^2)$ are the electromagnetic form factors of the neutrino,
corresponding to charge radius, magnetic moment (MM), electric dipole moment (EDM) and anapole moment (AM),
respectively, at $q^2=0$ \cite{Escribano,Vogel,Bernabeu1,Bernabeu2,Dvornikov,Giunti,Broggini}. The form factors
corresponding to charge radius and the anapole moment, are not considered in this paper.

In the SM minimally extended, the neutrino magnetic moment is given by $\mu_\nu\simeq 3.1\times 10^{-19}(m_{\nu_i}/1 \hspace{1mm} eV)\mu_B$ \cite{Fujikawa,Shrock}. Current limits on these magnetic moments are several orders of magnitude larger, so that a magnetic moment
close to these limits would indicate a window for probing effects induced by new physics beyond the SM \cite{Fukugita}.
Similarly, a neutrino electric dipole moment will also point to new physics and will be of relevance in astrophysics
and cosmology, as well as terrestrial neutrino experiments \cite{Cisneros}. In the case of the magnetic moment of the
$\nu_e$ the best bound is derived from globular cluster red giants energy loss \cite{Raffelt},

\begin{equation}
\mu_{\nu_e} < 3\times 10^{-12} \mu_B,
\end{equation}

\noindent is far from the SM value. From the experimental side of view, the best current laboratory bound

\begin{equation}
\mu_{\nu_e} < 2.9\times 10^{-11} \mu_B, \hspace{5mm} 90\%\hspace{0.8mm}C.L.,
\end{equation}

\noindent is obtained in $\bar\nu_e-e^-$ elastic scattering experiment GEMMA \cite{Bed}, which
is an order of magnitude larger than the constraint obtained in astrophysics \cite{Raffelt}.

For the magnetic moment of the muon-neutrino the current best limit has been obtained in
the LSND experiment \cite{Auerbach}

\begin{equation}
\mu_{\nu_\mu} \leq 6.8\times 10^{-10} \mu_B, \hspace{5mm} 90\%\hspace{0.8mm}C.L..
\end{equation}

In the case of the electric dipole moment $d_{\nu_e, \nu_\mu}$ \cite{Aguila} the best limits are:

\begin{equation}
d_{{\nu}_e, {\nu}_\mu} < 2\times 10^{-21} (e cm), \hspace{5mm} 95\%\hspace{0.8mm}C.L..\\
\end{equation}

Therefore, in general, it is worth investigating in deeper way tau-neutrino properties because their bounds are less restrictive.
These neutrinos correspond to the more massive third generation of leptons and possibly possess the largest mass and the largest
magnetic and electric dipole moments.

\subsection{$pp \to \nu_\tau \bar \nu_\tau \gamma+X$ cross section beyond the SM}

We study the process of simple production of photon in association with a pairs of massive neutrinos that could be observed
at the LHC, the schematic diagram is given in Fig. 1. The double production of $\nu_\tau$ may take place due the reaction

\begin{equation}
p p \to (\gamma, Z) \to \nu_\tau  \bar \nu_\tau  \gamma + X,
\end{equation}

\noindent where $X$ is a nondetecting hadron state. The Feynman diagrams of the subprocess

\begin{equation}
q\bar q \to (\gamma, Z) \to \nu_\tau  \bar \nu_\tau  \gamma,
\end{equation}

\noindent are shown in Fig. 2. The subprocess $q \bar q \to \nu_\tau \bar\nu_\tau \gamma$ is described by 8 tree-level Feynman
diagrams containing effective $\nu_\tau \bar\nu_\tau \gamma$ coupling. In calculations, we have taken into account subprocess
$q \bar q \to \nu_\tau \bar\nu_\tau \gamma$ $(q, \bar q=u, \bar u, d, \bar d, s, \bar s, c, \bar c)$. The $b$ quarks distribution
is not included in the calculations because its contribution is too small.

The analytical expression for the amplitude square is quite
lengthy so we do not present it here.  Instead, we present numerical fit functions for the total cross sections with respect to
center-of-mass energy and in terms of the form factors $F_2$ and $F_3$.

$\bullet$ For $\sqrt{s}=8\hspace{0.8mm} TeV$.

\begin{eqnarray}
\sigma(F_2)&=&(7.647 \times 10^{11}) F^4_2 + (56423) F^3_2 + (7486) F^2_2 - (0.033) F_2 + 0.0096,   \nonumber \\
\sigma(F_3)&=&(7.647 \times 10^{11}) F^4_3 + (7395) F^2_3 + 0.0096.
\end{eqnarray}

$\bullet$ For $\sqrt{s}=13\hspace{0.8mm} TeV$.

\begin{eqnarray}
\sigma(F_2)&=&(2.864 \times 10^{12}) F^4_2 - (2.343 \times 10^{-7}) F^3_2 + (19915) F^2_2 + 0.02,  \nonumber  \\\
\sigma(F_3)&=&(2.864 \times 10^{12}) F^4_3 + (19915) F_3^2 + 0.02.
\end{eqnarray}

$\bullet$ For $\sqrt{s}=14\hspace{0.8mm} TeV$.

\begin{eqnarray}
\sigma(F_2)&=&(3.463 \times 10^{12}) F^4_2 + (1.021 \times 10^{-7}) F^3_2 + (24582) F^2_2 + 0.022,   \nonumber  \\
\sigma(F_3)&=&(3.463 \times 10^{12}) F^4_3 + (24582) F^2_3 + 0.022.
\end{eqnarray}

$\bullet$ For $\sqrt{s}=33\hspace{0.8mm} TeV$.

\begin{eqnarray}
\sigma(F_2)&=&(2.645 \times 10^{13}) F^4_2 + (7.65 \times 10^{-6}) F^3_2 + (88985) F^2_2 - (2.577 \times 10^{-12}) F_2 + 0.0627,  \nonumber  \\
\sigma(F_3)&=&(2.645 \times 10^{13}) F^4_3 + (88985) F^2_3 + 0.0627.
\end{eqnarray}

\noindent In the expressions for the total cross section (8)-(11), the coefficients of $F_2 (F_3)$ given the anomalous contribution,
while the independent terms of $F_2 (F_3)$ correspond to the cross section at $F_2=F_3=0$ and represents the SM cross section magnitude.

\subsection{Bounds on the anomalous couplings at the LHC}

We have addressed a comprehensive study of the total cross section $\sigma_{Tot}=\sigma_{Tot}(\sqrt{s}, F_1, F_2)$ for
the channel of double tau-neutrino production in association with a photon $pp \to (\gamma, Z) \to \nu \bar \nu \gamma + X$.
This has been done as a function of the parameters of the reaction $pp \to (\gamma, Z) \to \nu \bar \nu \gamma + X$ as is
the MM and EDM, and of the parameters used by the LHC, that is $\sqrt{s}=8, 13, 14, 33\hspace{0.8mm}TeV$ and
${\cal L}=20, 50, 100,200, 500, 1000, 3000\hspace{0.8mm}fb^{-1}$. It should be mentioned that the fiducial phase space for
this measurement is defined by the requirements of photon transverse energy $E^\gamma_T$ and photon pseudorapidity $\eta^\gamma$.
The pseudorapidity requirement reduces the contamination from other particles misidentified as photons. In addition, for our study
we consider the CTEQ6L1 parton distribution functions (PDF) \cite{Jonathan} and we apply the following cuts to
reduce the background and to optimize the signal sensitivity:

\begin{eqnarray}
\begin{array}{c}
E^\gamma_T> 150\hspace{0.8mm}GeV,\\
p^{(\nu, \bar \nu)}_T >150\hspace{0.8mm}GeV,\\
|\eta^{\gamma}|< 2.37,
\end{array}
\end{eqnarray}

\noindent for the photon transverse energy $(E^\gamma_T)$, transverse momentum of the $Z$ boson decaying to a neutrino pair $(p^{(\nu, \bar \nu)}_T)$
and pseudorapidity $(\eta^\gamma)$ which are reported in Ref. \cite{Aad} by the ATLAS collaboration at the LHC. Furthermore, ours calculations are realized
via the computer program CALCHEP 3.6.30 \cite{Calhep}, which can computate the Feynman diagrams, integrate over multiparticle phase space and event
simulation at parton level.

The experimental systematic uncertainty on the $\nu_\tau\bar\nu_\tau\gamma$ cross section measurement is the uncertainty
in events with objects misidentified as photons. These events include jets fragmenting to photons and electrons.

The theoretical uncertainties that contribute to the extraction of the measured cross section arise from imprecise
knowledge of parton density function, from the choice of QCD scales, of the electroweak corrections and of the systematic
uncertainty related to the background from jets misidentified as photons. For a more detailed description on the uncertainties
we suggest the reader consult Ref. \cite{Aad}.

The systematic uncertainties for the collider, for the determination of the $\nu \bar \nu \gamma$ cross section, as well as for the magnetic moment and
the electric dipole moment for $E^\gamma_T$, $p_T^{(\nu, \bar \nu)}$ and $|\eta^\gamma|$ given in Eq. (12) is calculated
using the following formulae \cite{Sahin1,Billur,Ozguven,Koksal}:

\begin{equation}
\chi^2=\biggl(\frac{\sigma_{SM}-\sigma_{NP}(F_2, F_3)}{\sigma_{SM}\delta}\biggr)^2,
\end{equation}

\noindent where $\sigma_{NP}(F_2, F_3)$ is the total cross section including contributions from the SM
and new physics, $\delta=\sqrt{(\delta_{st})^2+(\delta_{sys})^2}$, $\delta_{st}=\frac{1}{\sqrt{N_{SM}}}$
is the statistical error, $\delta_{sys}$ is the systematic error and $N_{SM}$ is the number of signal expected
events $N_{SM}={\cal L}_{int}\times \sigma_{SM}$ where ${\cal L}_{int}$ is the integrated LHC luminosity.

\begin{table}[!ht]
\caption{Sensitivity on the $\mu_{\nu_\tau}$ magnetic moment and the $d_{\nu_\tau}$ electric dipole moment for $\sqrt{s}=8\hspace{0.8mm}TeV$
and ${\cal L}=20, 50, 100\hspace{0.8mm}fb^{-1}$ at $95\%$ C.L. through the process $pp \to \nu_\tau \bar\nu_\tau \gamma +X $.}
\begin{center}
 \begin{tabular}{ccccc}
\hline\hline
\multicolumn{5}{c}{ $95\%$ C.L.}\\
 \hline
 \cline{1-5} $\sqrt{s}\hspace{0.8mm}(TeV)$  & \hspace{1cm} ${\cal L}\hspace{0.8mm}(fb^{-1})$  & \hspace{1.5cm} $\delta_{sys}$ & \hspace{1.5cm}
 $\mu_{\nu_\tau}(\mu_B)\times 10^{-6}$   & \hspace{1.7cm} $|d_{\nu_\tau}(e cm)|$ \\
\hline
8                                         &\hspace{8mm}  20   &\hspace{1.2cm} $0\%$   &\hspace{1.2cm} [-6.384; 6.400]    & \hspace{1.5cm}
$1.051\times 10^{-16}$   \\
8                                         &\hspace{8mm}  20  &\hspace{1.2cm}  $5\%$   &\hspace{1.2cm} [-6.742; 6.756]    & \hspace{1.5cm}
$1.110\times 10^{-16}$   \\
8                                         &\hspace{8mm}  20  &\hspace{1.2cm}  $10\%$  &\hspace{1.2cm} [-7.402; 7.414]    & \hspace{1.5cm}
$1.218\times 10^{-16}$   \\
\hline
8                                         &\hspace{8mm}  50   &\hspace{1.2cm} $0\%$   &\hspace{1.2cm} [-5.606; 5.626]    & \hspace{1.5cm}
$9.243\times 10^{-17}$  \\
8                                         &\hspace{8mm}  50  &\hspace{1.2cm}  $5\%$   &\hspace{1.2cm} [-6.271; 6.287]    & \hspace{1.5cm}
$1.033\times 10^{-16}$  \\
8                                         &\hspace{8mm}  50  &\hspace{1.2cm}  $10\%$  &\hspace{1.2cm} [-7.173; 7.186]    & \hspace{1.5cm}
$1.181\times 10^{-16}$  \\
\hline
8                                         &\hspace{8mm}  100   &\hspace{1.2cm} $0\%$  &\hspace{1.2cm} [-5.068; 5.092]    & \hspace{1.5cm}
$8.362\times 10^{-17}$  \\
8                                         &\hspace{8mm}  100  &\hspace{1.2cm}  $5\%$  &\hspace{1.2cm} [-6.047; 6.064]    & \hspace{1.5cm}
$9.965\times 10^{-17}$  \\
8                                         &\hspace{8mm}  100  &\hspace{1.2cm}  $10\%$ &\hspace{1.2cm} [-7.085; 7.098]    & \hspace{1.5cm}
$1.166\times 10^{-16}$  \\
\hline\hline
\end{tabular}
\end{center}
\end{table}

\begin{table}[!ht]
\caption{Sensitivity on the $\mu_{\nu_\tau}$ magnetic moment and the $d_{\nu_\tau}$ electric dipole moment for $\sqrt{s}=13\hspace{0.8mm}TeV$
and ${\cal L}=20, 50, 100, 200\hspace{0.8mm}fb^{-1}$ at $95\%$ C.L. through the process $pp \to \nu_\tau \bar\nu_\tau \gamma +X $.}
\begin{center}
 \begin{tabular}{ccccc}
\hline\hline
\multicolumn{5}{c}{ $95\%$ C.L.}\\
 \hline
 \cline{1-5} $\sqrt{s}\hspace{0.8mm}(TeV)$  & \hspace{1cm} ${\cal L}\hspace{0.8mm}(fb^{-1})$  & \hspace{1.5cm} $\delta_{sys}$ & \hspace{1.5cm}
 $|\mu_{\nu_\tau}(\mu_B)\times 10^{-6}|$   & \hspace{1.7cm} $|d_{\nu_\tau}(e cm)|$ \\
\hline
13                                         &\hspace{8mm}  20   &\hspace{1.2cm} $0\%$   &\hspace{1.2cm} 4.994    & \hspace{1.5cm}
$8.221\times 10^{-17}$   \\
13                                         &\hspace{8mm}  20  &\hspace{1.2cm}  $5\%$   &\hspace{1.2cm} 5.504    & \hspace{1.5cm}
$9.064\times 10^{-17}$   \\
13                                         &\hspace{8mm}  20  &\hspace{1.2cm}  $10\%$  &\hspace{1.2cm} 6.243    & \hspace{1.5cm}
$1.028\times 10^{-16}$   \\
\hline
13                                         &\hspace{8mm}  50   &\hspace{1.2cm} $0\%$   &\hspace{1.2cm} 4.379    & \hspace{1.5cm}
$7.208\times 10^{-17}$  \\
13                                         &\hspace{8mm}  50  &\hspace{1.2cm}  $5\%$   &\hspace{1.2cm} 5.237    & \hspace{1.5cm}
$8.626\times 10^{-17}$  \\
13                                         &\hspace{8mm}  50  &\hspace{1.2cm}  $10\%$  &\hspace{1.2cm} 6.135    & \hspace{1.5cm}
$1.010\times 10^{-16}$  \\
\hline
13                                         &\hspace{8mm}  100   &\hspace{1.2cm} $0\%$  &\hspace{1.2cm} 3.954    & \hspace{1.5cm}
$6.508\times 10^{-17}$  \\
13                                         &\hspace{8mm}  100  &\hspace{1.2cm}  $5\%$  &\hspace{1.2cm} 5.125    & \hspace{1.5cm}
$8.442\times 10^{-17}$  \\
13                                         &\hspace{8mm}  100  &\hspace{1.2cm}  $10\%$ &\hspace{1.2cm} 6.096    & \hspace{1.5cm}
$1.004\times 10^{-16}$  \\
\hline
13                                         &\hspace{8mm}  200   &\hspace{1.2cm} $0\%$  &\hspace{1.2cm} 3.559    & \hspace{1.5cm}
$5.859\times 10^{-17}$  \\
13                                         &\hspace{8mm}  200  &\hspace{1.2cm}  $5\%$  &\hspace{1.2cm} 5.062    & \hspace{1.5cm}
$8.340\times 10^{-17}$  \\
13                                         &\hspace{8mm}  200  &\hspace{1.2cm}  $10\%$ &\hspace{1.2cm} 6.076    & \hspace{1.5cm}
$1.001\times 10^{-16}$  \\
\hline\hline
\end{tabular}
\end{center}
\end{table}

\begin{table}[!ht]
\caption{Sensitivity on the $\mu_{\nu_\tau}$ magnetic moment and the $d_{\nu_\tau}$ electric dipole moment for $\sqrt{s}=14\hspace{0.8mm}TeV$
and ${\cal L}=20, 50, 100, 200\hspace{0.8mm}fb^{-1}$ at $95\%$ C.L. through the process $pp \to \nu_\tau \bar\nu_\tau \gamma +X $.}
\begin{center}
 \begin{tabular}{ccccc}
\hline\hline
\multicolumn{5}{c}{ $95\%$ C.L.}\\
 \hline
 \cline{1-5} $\sqrt{s}\hspace{0.8mm}(TeV)$  & \hspace{1cm} ${\cal L}\hspace{0.8mm}(fb^{-1})$  & \hspace{1.5cm} $\delta_{sys}$ & \hspace{1.5cm}
 $|\mu_{\nu_\tau}(\mu_B)\times 10^{-6}|$   & \hspace{1.7cm} $|d_{\nu_\tau}(e cm)|$ \\
\hline
14                                         &\hspace{8mm}  20   &\hspace{1.2cm} $0\%$   &\hspace{1.2cm} 4.994    & \hspace{1.5cm}
$7.875\times 10^{-17}$   \\
14                                         &\hspace{8mm}  20  &\hspace{1.2cm}  $5\%$   &\hspace{1.2cm} 5.320    & \hspace{1.5cm}
$8.745\times 10^{-17}$   \\
14                                         &\hspace{8mm}  20  &\hspace{1.2cm}  $10\%$  &\hspace{1.2cm} 6.063    & \hspace{1.5cm}
$9.969\times 10^{-17}$   \\
\hline
14                                         &\hspace{8mm}  50   &\hspace{1.2cm} $0\%$   &\hspace{1.2cm} 4.379    & \hspace{1.5cm}
$6.894\times 10^{-17}$  \\
14                                         &\hspace{8mm}  50  &\hspace{1.2cm}  $5\%$   &\hspace{1.2cm} 5.074    & \hspace{1.5cm}
$8.343\times 10^{-17}$  \\
14                                         &\hspace{8mm}  50  &\hspace{1.2cm}  $10\%$  &\hspace{1.2cm} 5.966    & \hspace{1.5cm}
$9.809\times 10^{-17}$  \\
\hline
14                                         &\hspace{8mm}  100   &\hspace{1.2cm} $0\%$  &\hspace{1.2cm} 3.954    & \hspace{1.5cm}
$6.216\times 10^{-17}$  \\
13                                         &\hspace{8mm}  100  &\hspace{1.2cm}  $5\%$  &\hspace{1.2cm} 4.972    & \hspace{1.5cm}
$8.175\times 10^{-17}$  \\
14                                         &\hspace{8mm}  100  &\hspace{1.2cm}  $10\%$ &\hspace{1.2cm} 5.931    & \hspace{1.5cm}
$9.752\times 10^{-17}$  \\
\hline
14                                         &\hspace{8mm}  200   &\hspace{1.2cm} $0\%$  &\hspace{1.2cm} 3.559    & \hspace{1.5cm}
$5.586\times 10^{-17}$  \\
14                                         &\hspace{8mm}  200  &\hspace{1.2cm}  $5\%$  &\hspace{1.2cm} 4.916    & \hspace{1.5cm}
$8.083\times 10^{-17}$  \\
14                                         &\hspace{8mm}  200  &\hspace{1.2cm}  $10\%$ &\hspace{1.2cm} 5.913    & \hspace{1.5cm}
$9.722\times 10^{-17}$  \\
\hline\hline
\end{tabular}
\end{center}
\end{table}

\begin{table}[!ht]
\caption{Sensitivity on the $\mu_{\nu_\tau}$ magnetic moment and the $d_{\nu_\tau}$ electric dipole moment for $\sqrt{s}=33\hspace{0.8mm}TeV$
and ${\cal L}=100, 500, 1000, 3000\hspace{0.8mm}fb^{-1}$ at $95\%$ C.L. through the process $pp \to \nu_\tau \bar\nu_\tau \gamma +X $.}
\begin{center}
 \begin{tabular}{ccccc}
\hline\hline
\multicolumn{5}{c}{ $95\%$ C.L.}\\
 \hline
 \cline{1-5} $\sqrt{s}\hspace{0.8mm}(TeV)$  & \hspace{1cm} ${\cal L}\hspace{0.8mm}(fb^{-1})$  & \hspace{1.5cm} $\delta_{sys}$ & \hspace{1.5cm}
 $|\mu_{\nu_\tau}(\mu_B)\times 10^{-6}|$   & \hspace{1.7cm} $|d_{\nu_\tau}(e cm)|$ \\
\hline
33                                         &\hspace{8mm}  100   &\hspace{1.2cm} $0\%$   &\hspace{1.2cm} 2.009    & \hspace{1.5cm}
$4.256\times 10^{-17}$   \\
33                                         &\hspace{8mm}  100  &\hspace{1.2cm}  $5\%$   &\hspace{1.2cm} 3.887    & \hspace{1.5cm}
$6.391\times 10^{-17}$   \\
33                                         &\hspace{8mm}  100  &\hspace{1.2cm}  $10\%$  &\hspace{1.2cm} 4.667    & \hspace{1.5cm}
$7.674\times 10^{-17}$   \\
\hline
33                                         &\hspace{8mm}  500   &\hspace{1.2cm} $0\%$   &\hspace{1.2cm} 2.589    & \hspace{1.5cm}
$3.303\times 10^{-17}$  \\
33                                         &\hspace{8mm}  500  &\hspace{1.2cm}  $5\%$   &\hspace{1.2cm} 3.860    & \hspace{1.5cm}
$6.347\times 10^{-17}$  \\
33                                         &\hspace{8mm}  500  &\hspace{1.2cm}  $10\%$  &\hspace{1.2cm} 4.659    & \hspace{1.5cm}
$7.661\times 10^{-17}$  \\
\hline
33                                         &\hspace{8mm}  1000   &\hspace{1.2cm} $0\%$  &\hspace{1.2cm} 1.789    & \hspace{1.5cm}
$2.942\times 10^{-17}$  \\
33                                         &\hspace{8mm}  1000  &\hspace{1.2cm}  $5\%$  &\hspace{1.2cm} 3.857    & \hspace{1.5cm}
$6.342\times 10^{-17}$  \\
33                                         &\hspace{8mm}  1000  &\hspace{1.2cm}  $10\%$ &\hspace{1.2cm} 4.658    & \hspace{1.5cm}
$7.659\times 10^{-17}$  \\
\hline
33                                         &\hspace{8mm}  3000   &\hspace{1.2cm} $0\%$  &\hspace{1.2cm} 1.474    & \hspace{1.5cm}
$2.424\times 10^{-17}$  \\
33                                         &\hspace{8mm}  3000  &\hspace{1.2cm}  $5\%$  &\hspace{1.2cm} 3.855    & \hspace{1.5cm}
$6.338\times 10^{-17}$  \\
33                                         &\hspace{8mm}  3000  &\hspace{1.2cm}  $10\%$ &\hspace{1.2cm} 4.657    & \hspace{1.5cm}
$7.658\times 10^{-17}$  \\
\hline\hline
\end{tabular}
\end{center}
\end{table}

The total cross sections are presented as a function of $F_2$ or $F_3$ in Fig. 3 and Fig. 4 for the center-of-mass energies of
$\sqrt{s}= 8, 13, 14, 33\hspace{0.8mm}TeV$, respectively. The total cross sections $\sigma_{pp \to \nu_\tau \bar\nu_\tau \gamma+X}(\sqrt{s}, F_2, F_3)$
clearly show a strong dependence on the anomalous form factors $F_2$ and $F_3$, as well
as with the center-of-mass energy of the collider $\sqrt{s}$.

The total cross sections of $\sigma(pp \to \nu_\tau \bar\nu_\tau \gamma+X)$ as function of the form factors $F_2$ and $F_3$
are shown in Figs. 5-8. The total cross sections display a clear dependence on the form factors of the tau-neutrino.
From Eqs. 8-11 it can be seen that electric dipole moment terms are proportional to even powers due to this term can cause
CP violation. That is why the magnitudes of negative and positive parts of the bounds on the electric dipole moment are the
same. From these equations it can be show that a significant contribution from the odd powers of magnetic moments comes only
for $8\hspace{0.8mm}TeV$. This fact is reason of the little asymmetry in magnetic dipole moment bounds in Table II and symmetry
in others tables. This result can be explained as follows: interference terms between magnetic dipole moment and SM are small
due to neutrino mass and causes little difference for $8\hspace{0.8mm}TeV$ center-of-mass energy especially. Increasing energies
this difference is very small and can not be discerned from cross section values.

The allowed ranges for the anomalous magnetic moment $\mu_{\nu_\tau}$ and the electric dipole moment $d_{\nu_\tau}$ for the first, second,
third  and future runs of the LHC, that is $\sqrt{s}= 8, 13, 14, 33\hspace{0.8mm}TeV$, ${\cal L}=20, 50, 100, 200, 500, 1000, 3000\hspace{0.8mm}fb^{-1}$
at $95\%$ C.L. and we are considering systematic uncertainties of $\delta_{sys}=0, 5, 10\hspace{1mm}\%$ are shown in Tables II-V for $\nu_\tau \bar\nu_\tau \gamma$ vertex. We use the systematic uncertainties of $\delta_{sys}=0\%, 5\%, 10\%$ because there are no studies related to the systematic uncertainties for the process $pp \to \nu_\tau \bar\nu_\tau \gamma+X$ at the LHC. These results are compared in Table I with the previous Borexino, E872, CERN-WA-066 and L3 results \cite{Borexino,DONUT,A.M.Cooper,L3}. Although our studies have less sensitivity on the anomalous magnetic moment than for Borexino, however with increase of energy and luminosity at the LHC our bounds are competitive with those of E872 and CERN-WA-066, and furthermore, we obtain better bounds in comparison with those obtained at $90\%$ C.L. by the L3 collaboration at the Large Electron-Positron (LEP) collider.

The $95\%$ C.L. limits on each $\mu_{\nu_\tau}$ and $d_{\nu_\tau}$ parameters are obtained taken the $\mu_{\nu_\tau} (d_{\nu_\tau})$ one at the time.
The obtained limits are almost a factor of two better than the limit of the L3 collaboration. The best bounds obtained on $\mu_{\nu_\tau}$ and $d_{\nu_\tau}$
are $1.474\times 10^{-6}\hspace{0.8mm}\mu_B$ and $2.424\times 10^{-17}\hspace{0.8mm} ecm$, respectively, as shown in Tables II-V.
These limits are the most stringent to date, which are obtained through process $pp \to \nu_\tau \bar\nu_\tau \gamma+X$ and with the parameters of the LHC.

The $95\%$ C.L. regions in two-parameter space are shown as contours on the $(F_3-F_2)$ plane for $\sqrt{s}=8\hspace{0.8mm}TeV$,
${\cal L}=10, 50, 100 \hspace{0.8mm}fb^{-1}$, $\sqrt{s}=13\hspace{0.8mm}TeV$, ${\cal L}=20, 100, 200 \hspace{0.8mm}fb^{-1}$,
$\sqrt{s}=14\hspace{0.8mm}TeV$, ${\cal L}=20, 100, 200 \hspace{0.8mm}fb^{-1}$ and for $\sqrt{s}=33\hspace{0.8mm}TeV$,
${\cal L}=100, 500, 3000 \hspace{0.8mm}fb^{-1}$ in Figs. 9-12.

\section{Summary}

In this paper we have presented a calculation of the reaction $pp \to (\gamma, Z) \to \nu_\tau \bar\nu_\tau \gamma+X$ for general
$\nu_\tau \bar\nu_\tau \gamma$ anomalous coupling. Furthermore, we have study the possibility that the reaction
$pp \to (\gamma, Z) \to \nu_\tau \bar\nu_\tau \gamma+X$ can be used for probing the anomalous magnetic moment and the electric dipole
moment of the tau-neutrino  at the LHC with good precise.

We find that the total cross section $\sigma(pp \to \nu_\tau \bar\nu_\tau \gamma+X)$, as well as the dipole moments are very sensitive
to the parameters of the collider $\sqrt{s}$ and ${\cal L}$ as shown in Figs. 3-12 and in Tables II-V. In addition, for our study, we
consider cuts on the $E_T$, $p_T$ and $\eta$ as shown in Eq. (12) to improve the sensitivity on the cross section and on the dipole moments.
It is appropriate to mention that another interesting mechanism for studying the electromagnetic properties of the $\nu_\tau$, is the process
of production of a pair of neutrinos in association with two photons through the process $pp \to (\gamma, Z) \to \nu_\tau \bar\nu_\tau \gamma\gamma+X$.

In conclusion, we have study the possible manifestation of the MM and the EDM of the tau-neutrino in collisions $pp$ using the ATLAS
detector at the LHC. The channel $pp \to (\gamma, Z) \to \nu_\tau \bar\nu_\tau \gamma+X$ yields the most stringent limits on the $\mu_{\nu_\tau}$
and $d_{\nu_\tau}$ set at a hadron collider to date: $\mu_{\nu_\tau}=1.474\times 10^{-6}\hspace{0.8mm}\mu_B$ and $d_{\nu_\tau}=2.424\times 10^{-17}\hspace{0.8mm} ecm$. This is roughly a factor of 2.23 improvement over the result published in Ref. \cite{L3}.

\vspace*{2cm}

\begin{center}
{\bf Acknowledgments}
\end{center}%

The authors acknowledges Prof. Dimitrii Krasnopevtsev for valuable discussions and comments on the scattering cross section.
A. G. R. acknowledges support from SNI and PROFOCIE (M\'exico).

\pagebreak

\begin{figure}[t]
\centerline{\scalebox{0.75}{\includegraphics{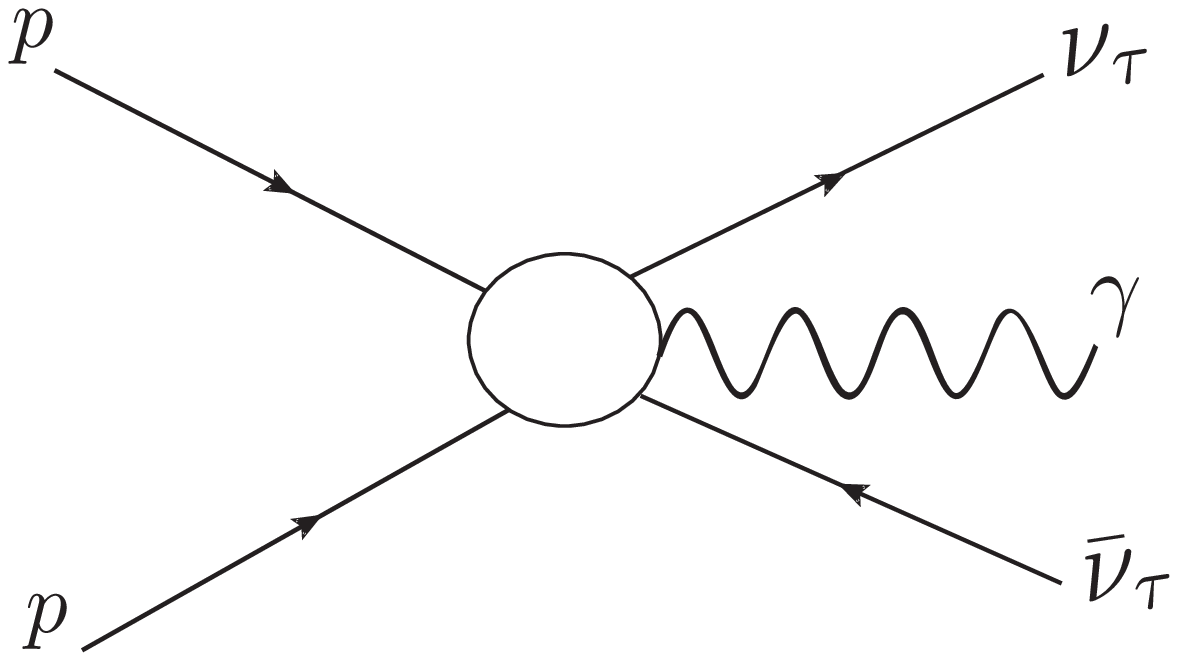}}}
\caption{ \label{fig:gamma1} A schematic diagram for the process $pp\rightarrow \nu_\tau \bar \nu_\tau \gamma+X$.}
\label{Fig.1}
\end{figure}

\begin{figure}[t]
\centerline{\scalebox{0.75}{\includegraphics{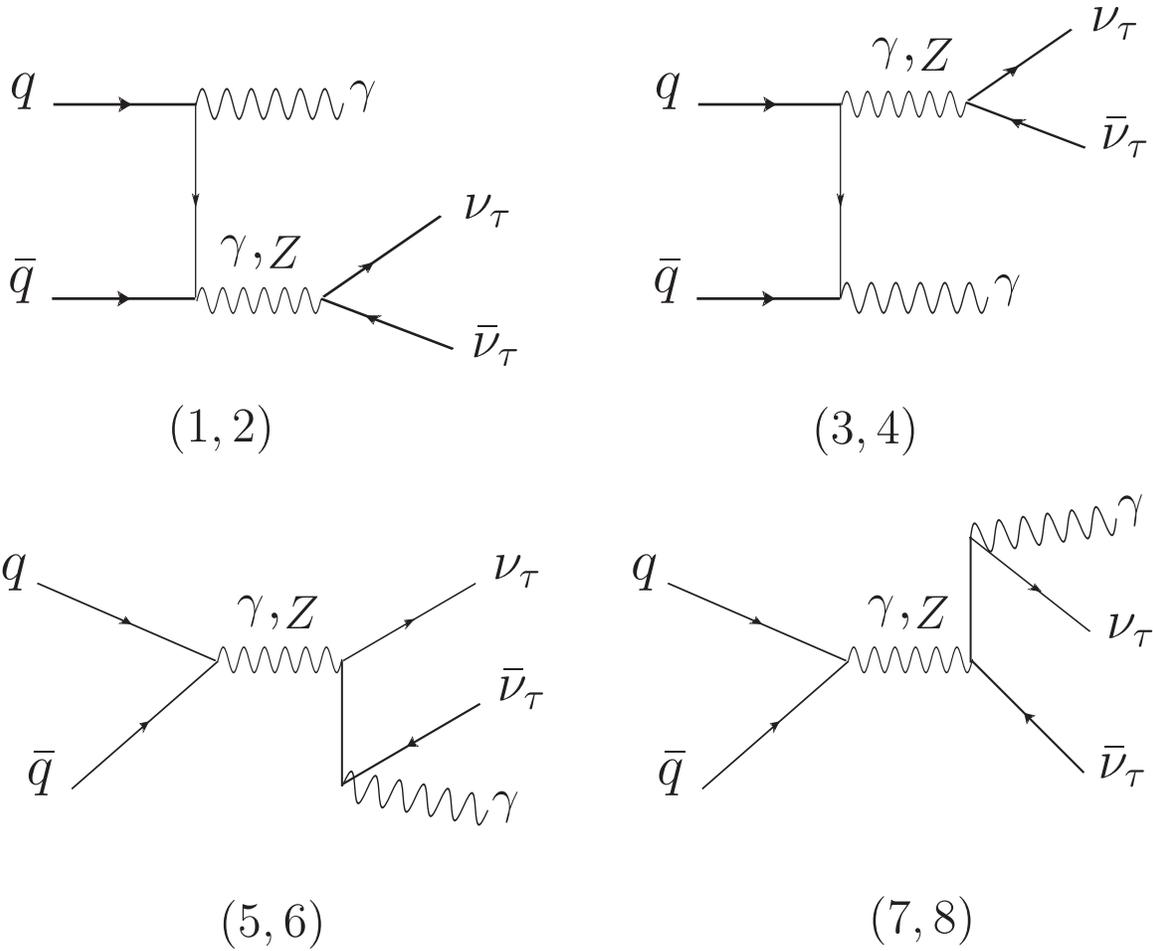}}}
\caption{ \label{fig:gamma2} The Feynman diagrams for the subprocesses
$q\bar q\rightarrow (\gamma, Z) \to \nu_\tau \bar \nu_\tau \gamma$.}
\label{Fig.2}
\end{figure}

\begin{figure}[t]
\centerline{\scalebox{1.5}{\includegraphics{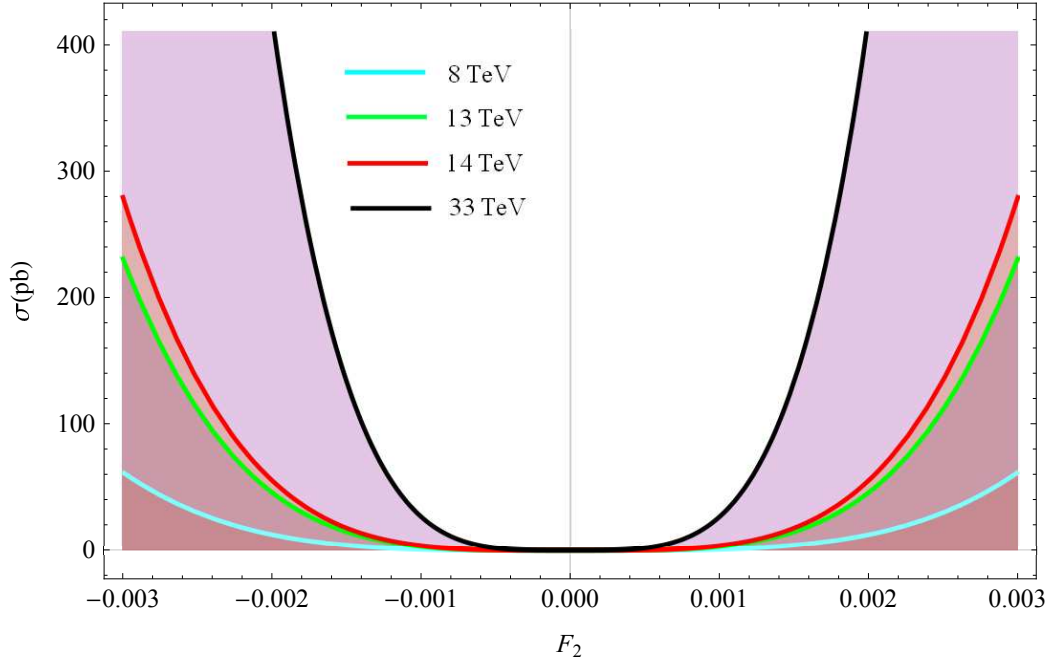}}}
\caption{ \label{fig:gamma1} The total cross sections of the process
$pp\rightarrow \nu_\tau \bar \nu_\tau \gamma+X$ as a function of $F_2$
for center-of-mass energies of $\sqrt{s}=8, 13, 14, 33$\hspace{0.8mm}$TeV$.}
\label{Fig.3}
\end{figure}

\begin{figure}[t]
\centerline{\scalebox{1.5}{\includegraphics{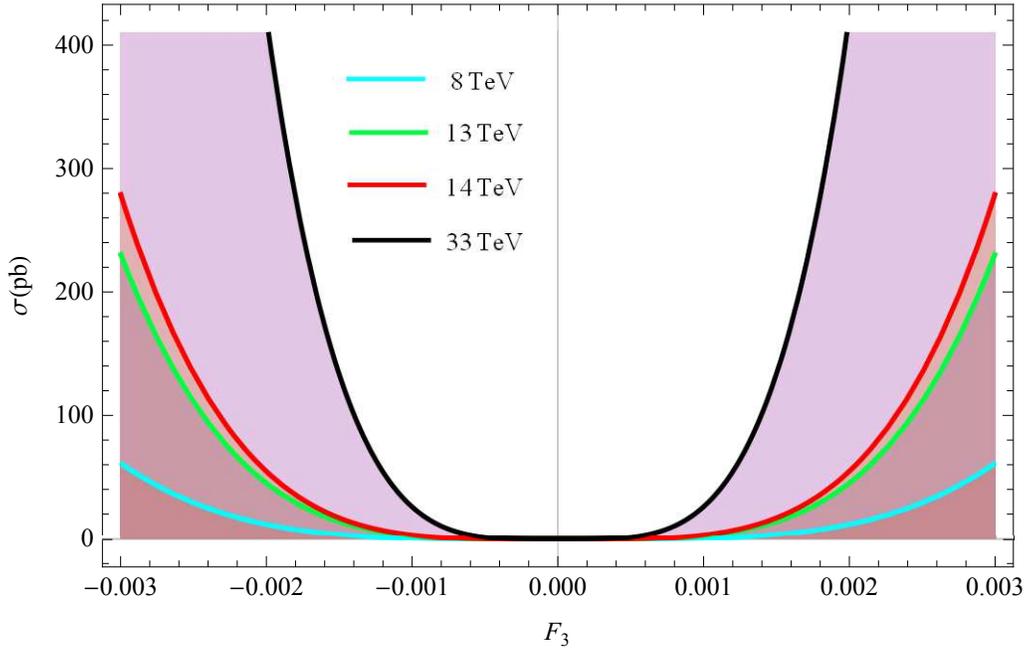}}}
\caption{ \label{fig:gamma2} Same as in Fig. 3, but for $F_3$.}
\label{Fig.4}
\end{figure}

\begin{figure}[t]
\centerline{\scalebox{1.2}{\includegraphics{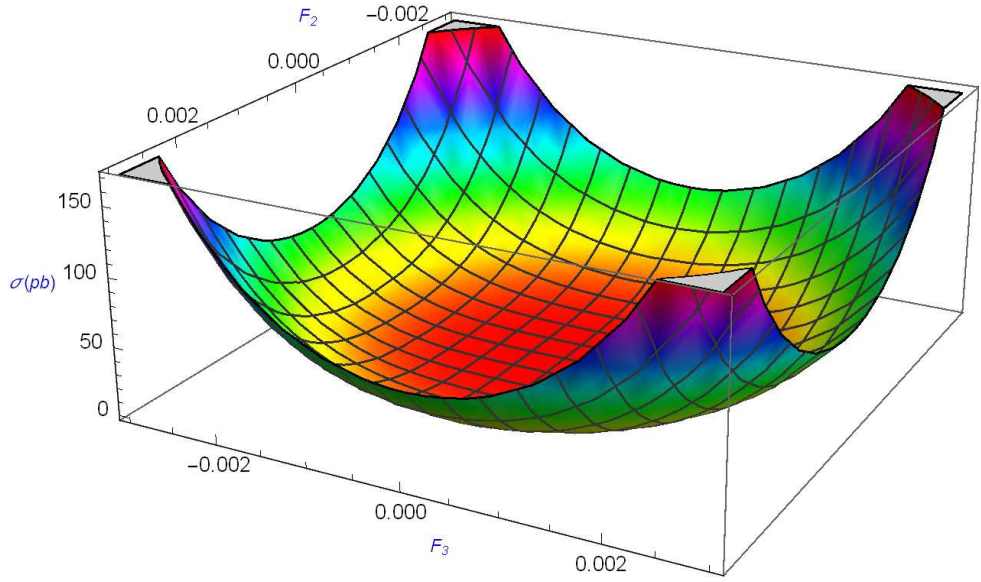}}}
\caption{ \label{fig:gamma3} The total cross sections of the process
$pp\rightarrow \nu_\tau \bar \nu_\tau \gamma+X$ as a function of $F_2$ and $F_3$
for center-of-mass energy of $\sqrt{s}=8\hspace{0.8mm}TeV$.}
\label{Fig.5}
\end{figure}

\begin{figure}[t]
\centerline{\scalebox{1.5}{\includegraphics{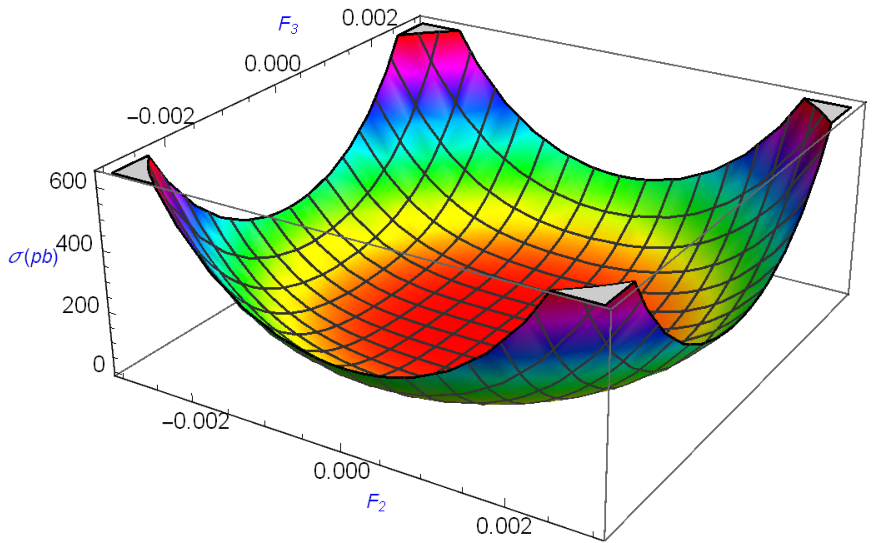}}}
\caption{ \label{fig:gamma15} Same as in Fig. 5, but for $\sqrt{s}=13\hspace{0.8mm}TeV$.}
\label{Fig.6}
\end{figure}

\begin{figure}[t]
\centerline{\scalebox{1.25}{\includegraphics{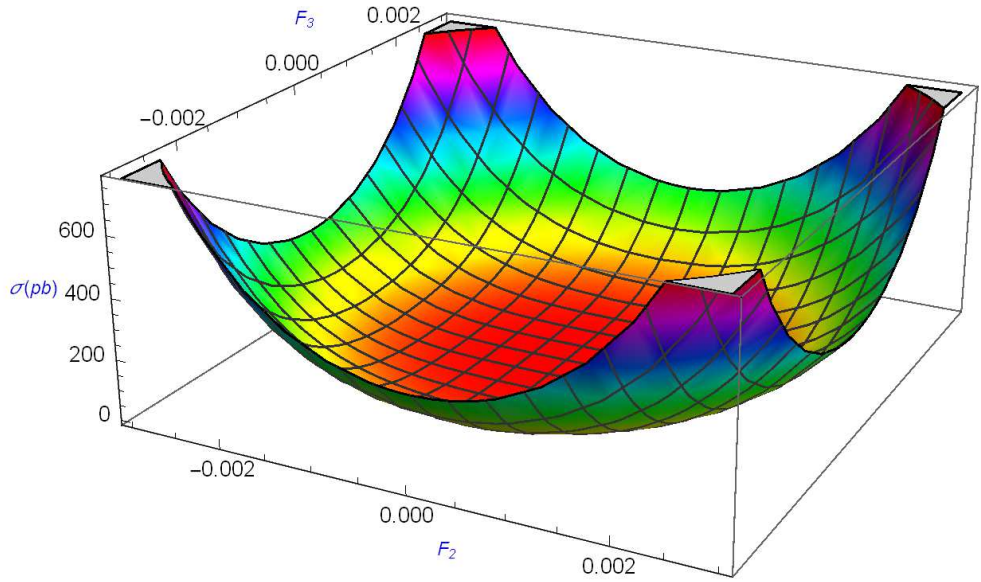}}}
\caption{ \label{fig:gamma15} Same as in Fig. 5, but for $\sqrt{s}=14\hspace{0.8mm}TeV$.}
\label{Fig.6}
\end{figure}

\begin{figure}[t]
\centerline{\scalebox{1.2}{\includegraphics{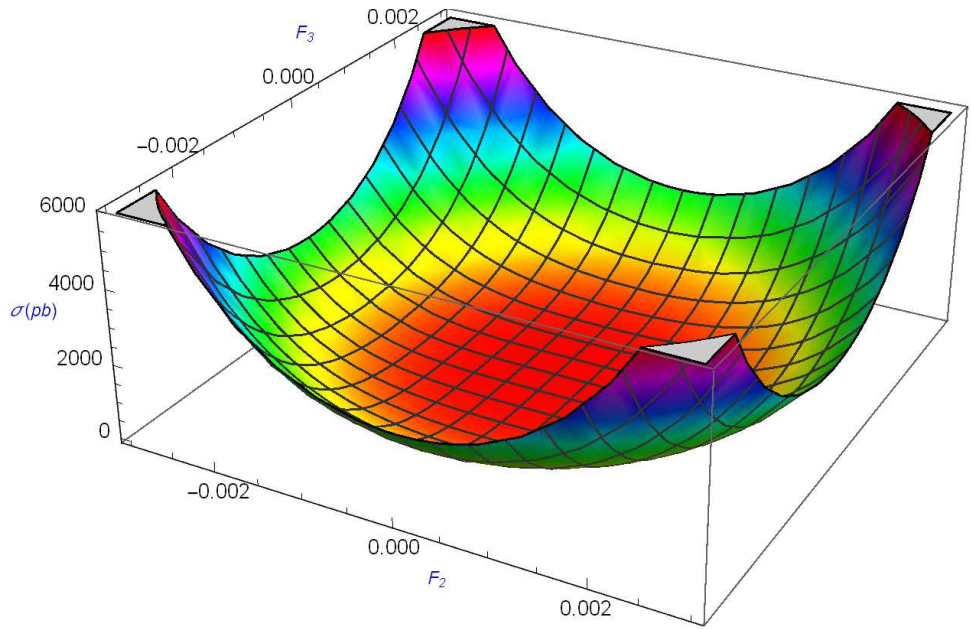}}}
\caption{ \label{fig:gamma15} Same as in Fig. 5, but for $\sqrt{s}=33\hspace{0.8mm}TeV$.}
\label{Fig.6}
\end{figure}

\begin{figure}[t]
\centerline{\scalebox{1.2}{\includegraphics{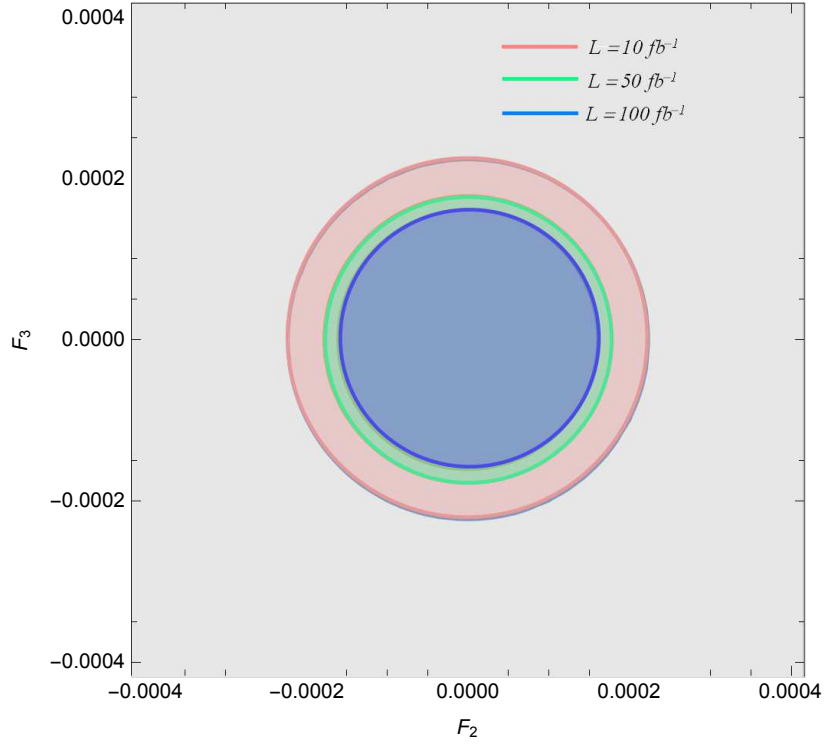}}}
\caption{ \label{fig:gamma15} Bounds contours at the $95\% \hspace{1mm}C.L.$ in the
$F_3-F_2$ plane for the process $pp\rightarrow \nu_\tau \bar \nu_\tau \gamma+X$
with the $\delta _{sys}=0\%$ and for center-of-mass energy of $\sqrt{s}=8\hspace{0.8mm}TeV$.}
\label{Fig.6}
\end{figure}

\begin{figure}[t]
\centerline{\scalebox{1.2}{\includegraphics{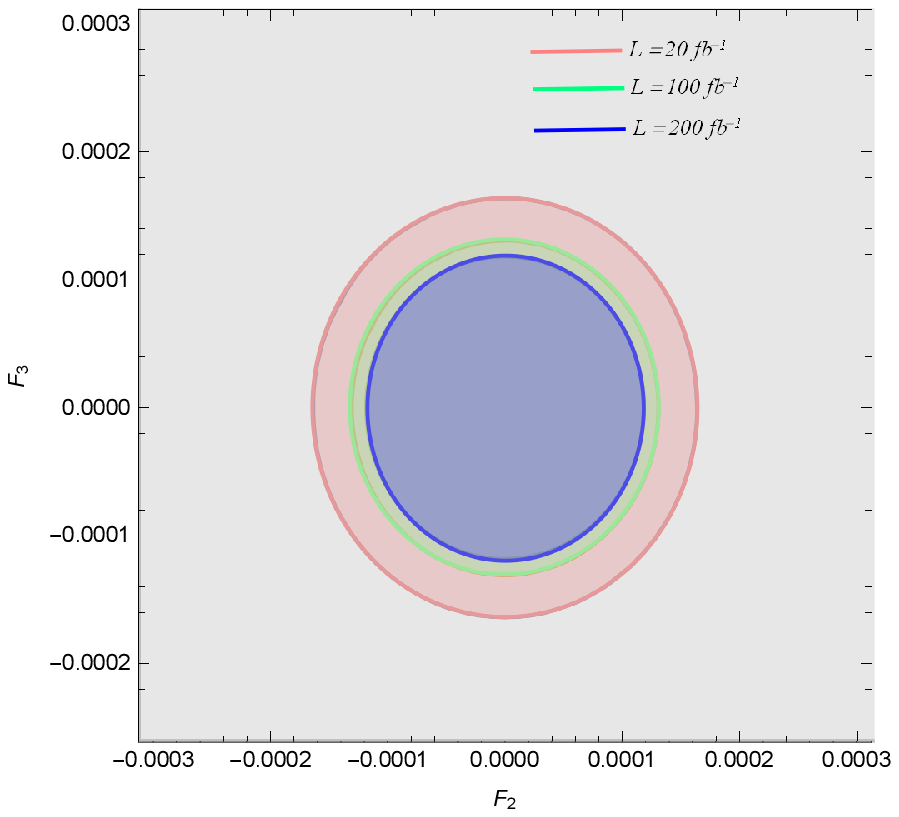}}}
\caption{ \label{fig:gamma15} Same as in Fig. 9, but for $\sqrt{s}=13\hspace{0.8mm}TeV$.}
\label{Fig.6}
\end{figure}

\begin{figure}[t]
\centerline{\scalebox{1.2}{\includegraphics{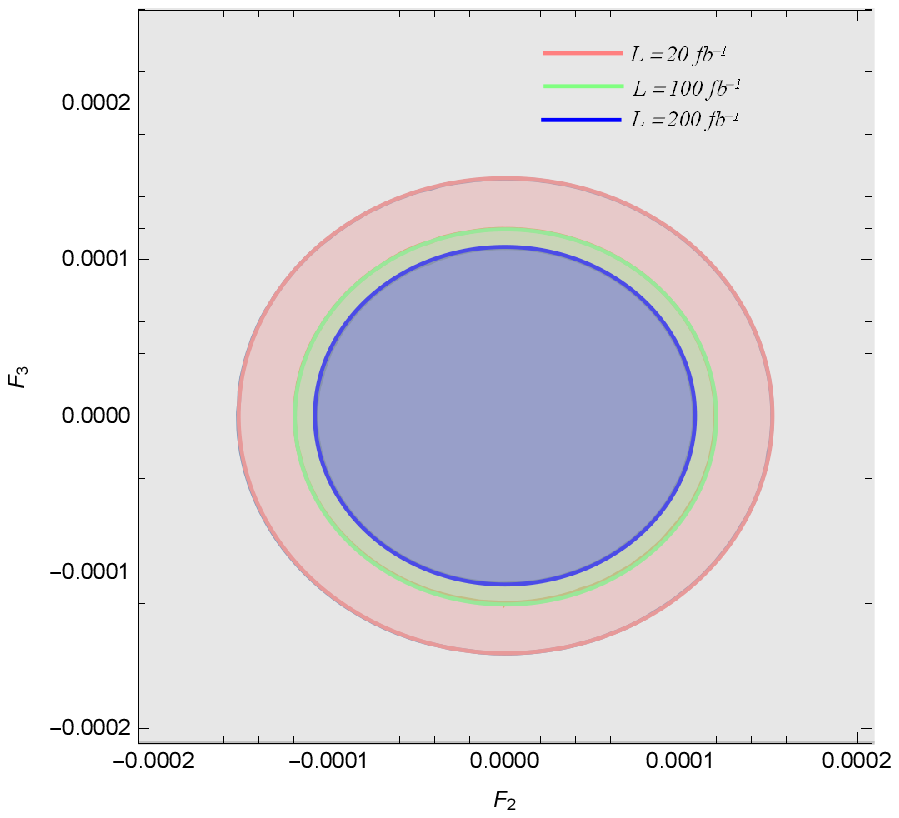}}}
\caption{ \label{fig:gamma15} Same as in Fig. 9, but for $\sqrt{s}=14\hspace{0.8mm}TeV$.}
\label{Fig.6}
\end{figure}

\begin{figure}[t]
\centerline{\scalebox{1.2}{\includegraphics{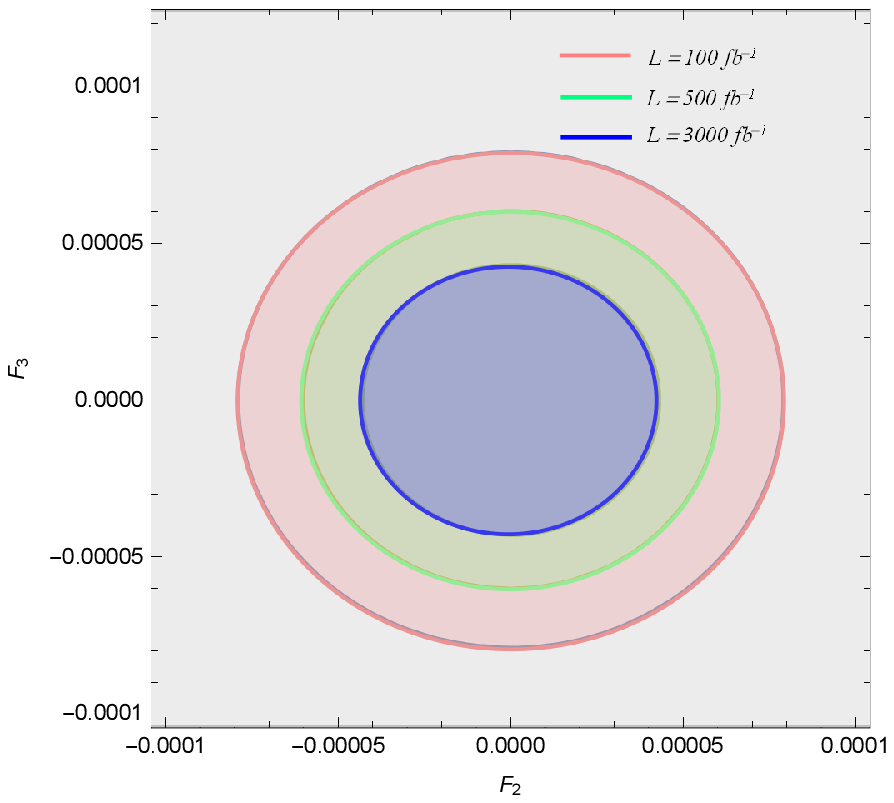}}}
\caption{ \label{fig:gamma15} Same as in Fig. 9, but for $\sqrt{s}=33\hspace{0.8mm}TeV$.}
\label{Fig.6}
\end{figure}

\end{document}